\documentclass[12pt]{iopart}

\usepackage[normalem]{ulem}
\usepackage{graphicx}
\usepackage{bm}
\usepackage[caption=false]{subfig}
\usepackage[hidelinks]{hyperref}
\hypersetup{colorlinks=true, linkcolor=blue, filecolor=magenta, urlcolor=cyan, citecolor=red}

\begin{document}

\title[Improved Band Gaps and Structural Properties from WFLSIC for Periodic Systems]{Improved Band Gaps and Structural Properties from Wannier-Fermi-L{\"o}wdin Self-Interaction Corrections for Periodic Systems}

\author{Ravindra Shinde, Sharma S. R. K. C. Yamijala, and Bryan M. Wong$^*$}

\address{ Department of Chemical \& Environmental Engineering, Materials Science \& Engineering Program, Department of Chemistry, and Department of Physics \& Astronomy, University of California-Riverside, Riverside, CA, USA.}
\ead{bryan.wong@ucr.edu}
\vspace{10pt}
\begin{indented}
\item[]\today
\end{indented}

\begin{abstract}
The accurate prediction of band gaps and structural properties in periodic systems continues to be one of the central goals of electronic structure theory. However, band gaps obtained from popular exchange-correlation functionals (such as LDA and PBE) are severely underestimated partly due to the spurious self-interaction error (SIE) inherent to these functionals. In this work, we present a new formulation and implementation of Wannier function-derived Fermi-L\"{o}wdin (WFL) orbitals for correcting the SIE in periodic systems. Since our approach utilizes a variational minimization of the self-interaction energy with respect to the Wannier charge centers, it is computationally more efficient than the HSE hybrid functional and other self-interaction corrections that require a large number of transformation matrix elements. Calculations on several (17 in total) prototypical molecular solids, semiconductors, and wide-bandgap materials show that our WFL self-interaction correction approach gives better band gaps and bulk moduli compared to semilocal functionals, largely due to the partial removal of self-interaction errors. 
\end{abstract}

%
%
%
%
%

\section{\label{sec:introduction} Introduction}
Kohn-Sham density functional theory (DFT)\cite{kohn-sham, hohenberg-kohn} is extensively used for predicting the electronic and structural properties of a variety of chemical/material systems. In this formally exact approach, the total energy of a many-electron system is a functional of the non-interacting electron density, which is given by
$E^\mathrm{KS} = T_s + E_\mathrm{H}[\rho_{\alpha}+\rho_{\beta}] + E_\mathrm{ext} + E_\mathrm{XC}[\rho_{\alpha},\rho_{\beta}]$, where $\rho_{\alpha}$ ($\rho_{\beta}$) is the electronic density for spin $\alpha$ ($\beta$). In the previous expression, $T_s$ is the kinetic energy of the fictitious non-interacting orbitals, $E_\mathrm{H}$ is the Hartree energy, and $E_\mathrm{ext}$ is the interaction energy due to an external potential (such as the nuclear attraction energy). The last $E_\mathrm{XC}$ term is the unknown exchange-correlation (XC) energy, which is often approximated using local or semi-local density functionals. Over the past several decades, these approximations have provided a useful balance between computational cost and accuracy.\cite{primer-dft}

Although the Kohn-Sham formalism has been used for a variety of chemical/material systems, it suffers from several issues: the XC potential decays too fast at asymptotic internuclear distances, the total energy of the system varies nonlinearly as a function of fractional occupation numbers, the band gaps of periodic systems are underestimated, and unphysical fractional charges appear for stretched internuclear distances (to name a few).\cite{primer-dft,Yang2017,koopman-marzari} There have been ongoing attempts to obtain better approximations for these XC functionals; however, the inaccuracy of all these Kohn-Sham DFT approaches can be traced to their inherent self-interaction error, which we describe further below.\cite{Zunger1980}

For a one-electron hydrogen atom, the total energy should not have any contributions from electron-electron repulsions, i.e., the $E_\mathrm{H}$ and $E_\mathrm{XC}$ energies should exactly cancel each other: $E_\mathrm{H}[\rho_{\alpha}]+E_\mathrm{XC}[\rho_{\alpha},0] = 0$. 
However, the local density (LDA) and generalized gradient density (GGA) approximations to the XC energy fail to satisfy this condition.  This spurious interaction of an electron with itself is known as the self-interaction error (SIE).

The SIE in many-electron systems is even more severe, and LDA/GGA functionals produce incorrect band gaps (among other incorrect electronic properties) when these approximations are invoked. To remove the SIE in a systematic way, Perdew and Zunger (PZ)\cite{Zunger1980} introduced a self-interaction correction (SIC) to the exchange-correlation energy ($E_\mathrm{XC}$). The corrected energy, $E_\mathrm{XC}^\mathrm{SIC}$, is defined as: 
\begin{equation}
E_\mathrm{XC}^\mathrm{SIC} = E_\mathrm{XC}^{\mathrm{approx}}[\rho_{\alpha},\rho_{\beta}] - 
\sum_{i \sigma} \left(
E_\mathrm{H}[\rho_{i\sigma}] + E_\mathrm{XC}^{\mathrm{approx}}[\rho_{i\sigma},0] \right)
\label{eqn:pz}
\end{equation}
where $E_\mathrm{XC}^{\mathrm{approx}}[\rho_{\alpha},\rho_{\beta}]$ is an approximate XC energy (i.e., from LDA or GGA), and the summation over orbitals and spins denotes the self-interaction energy contribution from each electron in orbital $i$ with spin $\sigma$ with orbital density $\rho_{i\sigma}$ (i.e., $E_\mathrm{H}[\rho_{i\sigma}]$ is the self-Coulomb part, and $E_\mathrm{XC}^{\mathrm{approx}}[\rho_{i\sigma},0]$ is the self-exchange-correlation part). 
The PZ-SIC approach has been widely used to obtain accurate total energies, ionization energies, and electron affinities of various finite systems.\cite{Zunger1980} Within the PZ-SIC approximation, the self-interaction correction is calculated using either canonical molecular orbitals (in general, delocalized) or localized orbitals. 

In recent work, Fermi-L\"{o}wdin orbitals were shown to more accurately correct for self-interaction errors compared to canonical orbitals in molecular systems.\cite{fermi-lowdin-fredy, fredy-ravindra-fractional-sic, fermi-lowdin-jcc, fermi-lowdin-jcp} However, in periodic systems, localized wavefunctions are required to compute the SIC using a direct implementation of the PZ formalism. Specifically, the PZ-SIC contribution is not invariant with respect to a unitary transformation of the occupied manifold and vanishes for extended Bloch wave functions.\cite{Stengel2008} Because of these limitations, previous researchers have suggested that localized functions are a more suitable choice to account for SIC in solids. \cite{sic-wannier-srep} 

To this end, Heaton et al. \cite{wannier-sic-HHL} and Stengel et al. \cite{Stengel2008} previously proposed that Wannier functions could be successfully used to compute the PZ-SIC of localized orbitals in solids. Using Wannier functions, Heaton et al.\cite{wannier-sic-HHL} considerably improved the LDA band gaps of solid argon and LiCl. In particular, their SIC Hamiltonian was not orbital dependent, and for a given $k$-point, the eigenvalues could be calculated using a single matrix diagonalization. Similarly, Stengel et al. used Wannier functions to compute self-interaction corrections to the LDA functional \cite{Stengel2008} and found that Wannier-function-based SIC tended to over-correct LDA band gaps. In particular, they also showed that SIC applications to transition metal oxides and elements with $d$-electrons were hindered by the breaking of spherical symmetry. 
To address transition metal oxide materials, a fully self-consistent, self-interaction corrected local spin density approach was developed by Svane and Gunnarsson to correct band gaps and magnetic moments.\cite{svane-prl} Similarly, Szotek et al. applied self-interaction corrections to the standard linear muffin-tin orbital model to substantially improve the band gaps of transition metal oxide materials.\cite{szotek-prb} Lastly, completely different approaches using self-interaction corrected pseudopotentials and exact exchange (EXX) were used by Vogel et al.\cite{SIC-PP-PRB} and Qteish et al.\cite{EXX-PRB}, respectively, to correct the LDA band gaps of group-III nitrides.

The shortcomings of these SIC methods motivated us to formulate and implement an alternative approach to calculating the SIC in periodic systems. In this work, we construct localized Fermi-L\"{o}wdin functions using Wannier functions for each band, which are then used to compute the Hartree and XC energy contribution. The Perdew-Zunger expression is then used to sum up the SIC, which is computed using localized Fermi-L\"{o}wdin functions for all the bands. We benchmark our implementation by computing the ionization potentials of a set of molecular systems and comparing them against an all-electron molecular FLO-SIC implementation as well as with experimentally available values.\cite{fredy-ravindra-fractional-sic} Using the self-interaction-corrected electronic wavefunctions and densities, we then calculate the bulk modulus and bandstructure of a few representative periodic systems. A comparison with available experimental values provides a validation of our results and useful guidelines for utilizing these Wannier-Fermi-L\"{o}wdin self-interaction corrections for periodic systems.

\section{\label{sec:methodology} Theory and Computational Details}
In the following subsections, we present derivations of Wannier-Fermi-L\"{o}wdin wavefunctions for periodic systems. Our derivation is then followed by the PZ expression for the calculation of the SIC energy and its minimization with respect to the Wannier charge centers. 

\subsection{Wannier-Fermi-L\"{o}wdin Orbital Method}
In a periodic system, the resultant Bloch states, $\psi_{n\bm{k}}(\bm{r})$, are characterized by a band index $n$ and a crystal momentum $\bm{k}$. We denote N$_b$ to be the total number of bands and $V$ the real-space unit cell volume. The generalized Bloch states $\psi_{n\bm{k}}(\bm{r})$ can be written in terms of the cell-periodic functions, $u_{n\bm{k}}(\bm{r})$:
\begin{equation}
 \psi_{n\bm{k}}(\bm{r}) = e^{i\bm{k.r}} u_{n\bm{k}}(\bm{r}).
\end{equation}
The cell-periodic function itself can be written in reciprocal space as
\begin{equation}
 \tilde{u}_{n\bm{k}}(\bm{G}) = \frac{1}{\sqrt{V}} \int_{cell} e^{-i\bm{G.r}} u_{n\bm{k}}(\bm{r}) d\bm{r}.
\end{equation}
Using the Bloch states, we can construct an orthogonal set of Wannier functions given by
\begin{equation}
 w_{0}(\bm{r}- \bm{R}) = \frac{V}{8\pi^{3}} \int_\mathrm{BZ} e^{i\bm{k.R}} \psi_{n\bm{k}}(\bm{r}) d\bm{k},
 \label{eqn:wannier-definition}
\end{equation}
where $V$ is the real-space unit cell volume, and the integral is carried out over the full Brillioin zone (BZ). If $\bm{R}=0$, Eqn. \ref{eqn:wannier-definition} can be interpreted as the Wannier function located in the ``home" unit cell. These Wannier functions are orthogonal to each other, and carry a gauge  freedom.\cite{Marzari2012}

Using the properties of Wannier functions, the general matrix elements of position operators between Wannier functions are given by
\begin{equation}
 \langle \bm{R} n | \bm{r} | \bm{0} m \rangle = i \frac{V}{8\pi^{3}} \int_\mathrm{BZ} e^{-i\bm{k.R}} \langle u_{n\bm{k}} | \nabla_{k} | u_{m\bm{k}} \rangle  d\bm{k},
\end{equation}
where $u_{n\bm{k}}(\bm{r})$ is the periodic part of the Bloch function.\cite{Marzari2012} 
In practice, the matrix elements in the above equation are not directly evaluated, rather the overlap between the Bloch orbitals are computed instead:\cite{Marzari2012}
\begin{equation}
 M_{m,n}^{\bm(k,b)} = \langle u_{m\bm{k}} | u_{n\bm{k+b}} \rangle.
\label{eqn:bloch-overlap}
 \end{equation}
This overlap matrix is used to compute the expectation value of the position operator $\langle \tilde{r_n} \rangle $, which takes the form:
\begin{equation}
 \langle \tilde{r_n} \rangle = -\frac{1}{N} \sum_{\bm{k,b}} w_{b} \bm{b} \mathrm{Im}\left[\mathrm{ln}\left(M_{n,n}^{\bm(k,b)}\right)\right].
 \label{eqn:wcc}
\end{equation}
Here, $\bm{b}$ is a vector connecting a $\bm{k}$ point to one of its neighbors, and $w_{b}$  is an appropriate geometric factor that depends on the number of points in the star stencil and its geometry.\cite{marzari-vanderbilt} We call these expectation values as Wannier charge centers (WCC), $\bm{a_{m \sigma}}$, for spin $\sigma$.

We can construct a set of transformed functions, derived from the Wannier functions, which will be used later to minimize the self-interaction energy. The first step consists of transforming the Wannier functions into Fermi functions, F$_{m\sigma} (\bm{r})$:

\begin{equation} 
 F_{m\sigma} (\bm{r}) = \frac{\sum_{n} w_{n \sigma}^* (\bm{a_{m \sigma}})  w_{n \sigma} (\bm{r}) }  {\sqrt{\sum_{n} |w_{n \sigma} (\bm{a_{m \sigma}}) |^2 }}.
\label{eqn:fermi-orbitals}
 \end{equation}
Since these transformed Fermi orbital functions are not generally orthogonal, a symmetric L\"{o}wdin orthogonalization procedure is invoked. This approach uses the eigenvectors and eigenvalues of the overlap matrix of Fermi orbital functions, where the Fermi orbital overlap, $S_{mn}^{\sigma}$, is given by
\begin{equation}
S_{mn}^{\sigma} = \langle F_{m \sigma} | F_{n \sigma} \rangle.
\label{eqn:fermi-overlap}
\end{equation}
Upon diagonalization of the Fermi orbital overlap matrix, we obtain the eigenvalues, $\lambda_{\alpha}^{\sigma}$. and corresponding eigenvectors, $T_{\alpha m}^{\sigma}$:
\begin{equation}
\sum_n S_{mn}^{\sigma} T_{\alpha n}^{\sigma}  = \lambda_{\alpha}^{\sigma}  T_{\alpha n}^{\sigma} .
\label{eqn:fermi-eigenvalue-equation}
\end{equation}
The Wannier-Fermi-L\"{o}wdin (WFL) functions are constructed using the matrix elements of the eigenvectors, which gives
\begin{equation}
| \phi_{l\sigma} \rangle = \sum_n  \phi_{ln}^{\sigma} | F_n^{\sigma} \rangle,
\quad \quad \mathrm{ with } \quad
\phi_{ln}^{\sigma} = \sum_{\alpha} \frac{T_{\alpha l}^{\sigma}T_{\alpha n}^{\sigma}}{\sqrt{\lambda_{\alpha}^{\sigma}}}.  \nonumber
\label{eqn:lowdin2}
\end{equation}

Densities are evaluated for each state using these WFL states and are used to compute the self-interaction corrections for that particular state. The SIC contribution for a given XC functional is calculated using the Perdew-Zunger expression for a given spin $\sigma$ and state $l$ using the expression
\begin{eqnarray}
E^{\mathrm{WFL-SIC}}_{\sigma} = -\sum_{l} \left\{
E_\mathrm{XC}[\rho_{l \sigma},0]  +  \frac{1}{2} \int \int d{\bf r}d{\bf r'}
\frac{\rho_{l \sigma}({\bf r})\rho_{l \sigma}({\bf r}')}
{|{\bf r}-{\bf r'}|} 
\right \},
\label{eqn:pz-sic}
\end{eqnarray}
where the first term, $E_\mathrm{XC}[\rho_{l \sigma},0]$, denotes the XC contribution, and the second term denotes the Coulomb contribution to the SIC energy calculated using the orbital charge density $\rho_{l\sigma}$. 

The WFL-SIC potential is obtained by evaluating the gradients of the SIC energy with respect to the WFL functions. The contribution from each orbital is added, and the WFL-SIC potential is obtained as
\begin{equation}
\hat{V}^{\mathrm{WFL-SIC}} = \sum_{n\sigma} \hat{V}^{\mathrm{WFL-SIC}}_{n\sigma} |\phi_{n\sigma}\rangle \langle \phi_{n\sigma}|.
\label{eqn:sic-potential}
\end{equation}
The self-interaction corrected energy given by Eqn. \ref{eqn:pz-sic} is then numerically minimized using a Powell minimization scheme with respect to the Wannier charge center (WCC) positions to obtain:
\begin{equation}
\frac{\partial E^{\mathrm{WFL-SIC}}_{\sigma}} {\partial \bm{a_{m \sigma}}} = 0 
\label{eqn:optimization}
\end{equation}
This condition gives rise to a unique set of WCCs in the neighbourhood of the initial WCCs. Although the initial Wannier functions are not maximally localized, this minimization implies a stable solution.\cite{fredy-ravindra-fractional-sic} 

The SIC energy minimization also implies that the anti-Hermitian component of the SIC potentials approaches zero:\cite{fermi-lowdin-fredy} 
\begin{equation}
\langle \phi_{m\sigma}| \hat{V}^{\mathrm{WFL-SIC}}_{n\sigma} - \hat{V}^{\mathrm{WFL-SIC}}_{m\sigma} |\phi_{n\sigma}\rangle  = 0.
\label{eqn:localization}
\end{equation}
The minimization of the SIC energy is carried out with respect to the 3$\times N_{\mathrm{occ}}$ WCC positions, which are far less than the $N_{\mathrm{occ}}\times N_{\mathrm{occ}}$ parameters used in full SIC calculations (the latter are used to minimize the energy with respect to the elements of the unitary transformation matrix of the occupied subspace).

\subsection{Computational Details and Implementation}
We implemented our Wannier function-based Fermi-L\"{o}wdin self-interaction formalism in the open-source GPAW package\cite{GPAW1}, which utilizes finite-difference real-space grids.\cite{GPAW2} The WFL-SIC implementation described in this work is specific to real-space grids; however, it can easily be extended to utilize plane-waves in the GPAW package. 

To benchmark and test our implementation, we  investigated a total of 17 different systems, which includes both periodic systems and molecules in a large periodic box. In all of our simulations, we used the PBE\cite{pbe1,pbe2} exchange-correlation functional with a 24$\times$24$\times$24 real-space grid to obtain a reference ground state, with the core electrons described with the projector augmented wave (PAW) method.\cite{paw-method} For the periodic solids, we sampled the Brillouin zone using a 3$\times$3$\times$3 Monkhorst--Pack grid.\cite{monkhorst-pack}  The molecular systems examined in this study were kept in a  6$\times$6$\times$6 \AA{} periodic box and were sampled at the $\Gamma{}$ point. Since our periodic solids were simulated with supercell structures (see the following paragraph for further details), the 3$\times$3$\times$3 $k$-grid was found to be sufficient to give converged results. The nuclei were relaxed until the forces on each atom were less than 0.01 eV/\AA{}. The Wannier functions were obtained using the formalism in Ref. \cite{part-wannier-gpaw}, and a $10^{-6}$ \AA{}$^2$ convergence criteria was used for minimizing the sum of the quadratic spreads of the Wannier functions about their centers of reference. Additionally, the resultant Wannier functions were orthogonalized again using a L\"{o}wdin symmetric orthogonalization. A numerical conjugate gradient method was used to minimize the SIC energy with respect to the Wannier charge center positions, and the energy was minimized until a $10^{-6}$ eV threshold was met.

Since the Wannierization module in GPAW does not support non-orthogonal cells, all of our systems were constructed with unit cells with lattice vectors that are orthogonal. Specifically for the cubic systems investigated in this work, these supercells (SCs) had a volume four times that of their non-orthogonal primitive cell. Hence, the resultant bandstructures of these SCs need to be unfolded back onto the primitive cell Brillouin zone (PCBZ), which were carried out using Popescu and Zunger's method, \cite{band_unfolding_zunger} as implemented in the GPAW package.\cite{GPAW2} The unfolded bandstructure can be represented by the spectral function:
\begin{equation}
A(\vec{k},\epsilon) = \sum_{m} P_{\vec{K}m}(\vec{k}) \delta \left(\epsilon_{{\vec{K}m}} - \epsilon \right), \label{eqn:spectral1}
\end{equation}
where the spectral weights, $P_{\vec{K}m}$, are defined by:
\begin{eqnarray}
P_{\vec{K}m}(\vec{k}) &=& \sum_{n} \left| \langle \phi_{\vec{K}m}^{\mathrm{SC}}  | \phi_{\vec{k}n}^{\mathrm{PC}} \rangle \right|^{2} \nonumber \\
                      &=& \sum_{\vec{g} \in \mathrm{PC}} \left| C_{\vec{K}m}^{\mathrm{SC}} \left(\vec{g} + \vec{k}  - \vec{K}\right)   \right|^{2}.
\label{eqn:spectral2}
\end{eqnarray}
Here, $\vec{k}$ and $\vec{K}$ are the wave vectors belonging to the PCBZ and SCBZ, respectively, and $\vec{g}$ belongs to the primitive cell reciprocal lattice vectors. For all of our WFL bandstructures and spectral functions, we multiplied the SIC potential with an arbitrary prefactor of $\frac{2}{3}$, which is a standard procedure to account for the overestimation of the SIC band gaps.\cite{fredy-ravindra-fractional-sic, losc-method} 

\section{\label{sec:level3} Results and Discussion}

\subsection{Molecular Systems}
To benchmark our WFL-SIC implementation, we calculated the ionization potentials (using IP$_\mathrm{calc}$ = --$\epsilon_\mathrm{HOMO}$) for a set of molecules and compared them against earlier experimental results and another independent FLO-SIC implementation. The latter is an all-electron approach implemented by us in the NWChem software package for computing the FLO-SIC total energy,\cite{fredy-ravindra-fractional-sic}, which accurately predicts molecular properties such as the total energy, atomization energy, ionization potential, and linearity with fractional occupation numbers. In Table \ref{tab:molecular} we list the ionization potentials obtained from both the FLO-SIC and WFL-SIC methods compared against available experimental values. From these results, we find that the ionization potentials computed using these two  completely different computational methods compare well with each other as well as with experimental benchmarks.

\begin{table}[h!]
\caption{\label{tab:molecular} 
 Experimental and calculated ionization potentials (IP$_\mathrm{calc}$ = --$\epsilon_\mathrm{HOMO}$) of various molecules. The FLO-SIC calculations were carried out with the all-electron NWChem software package using a PBE/cc-pVTZ reference state, and the WFL-SIC calculations utilized a real-space, finite difference grid-based approach with a PBE reference state.}
 \begin{indented}
\footnotesize
\item[]\begin{tabular*}{0.6\columnwidth}{@{}ccccc} 
\br
Sr.   &   System  &     FLO-SIC IP    &   WFL-SIC IP     &  Expt IP\\
No.   &           &     NWChem (eV)   &   GPAW (eV)      &  (eV)   \\
\mr
1     &   C$_2$H$_2$ &      11.94        &   11.53       &  11.40  \\
2     &   CO         &      14.92        &   14.68       &  14.10  \\
3     &   N$_2$      &      16.84        &   16.42       &  15.56  \\
4     &   H$_2$      &      16.74        &   16.63       &  15.42  \\
5     &   CH$_4$     &      15.60        &   15.32       &  12.60  \\
6     &   NH$_3$     &      12.01        &   11.95       &  10.07  \\
7     &   H$_2$O     &      14.01        &   14.10       &  12.62  \\
8     &   O$_2$      &      13.34        &   13.58       &  12.07  \\
9     &   CO$_2$     &      14.79        &   14.65       &  13.77  \\
\br
\end{tabular*}
\end{indented}
\end{table}
\normalsize

\subsection{Periodic Systems}


\begin{figure}[h!]
    \centering
    \includegraphics[width=0.75\columnwidth]{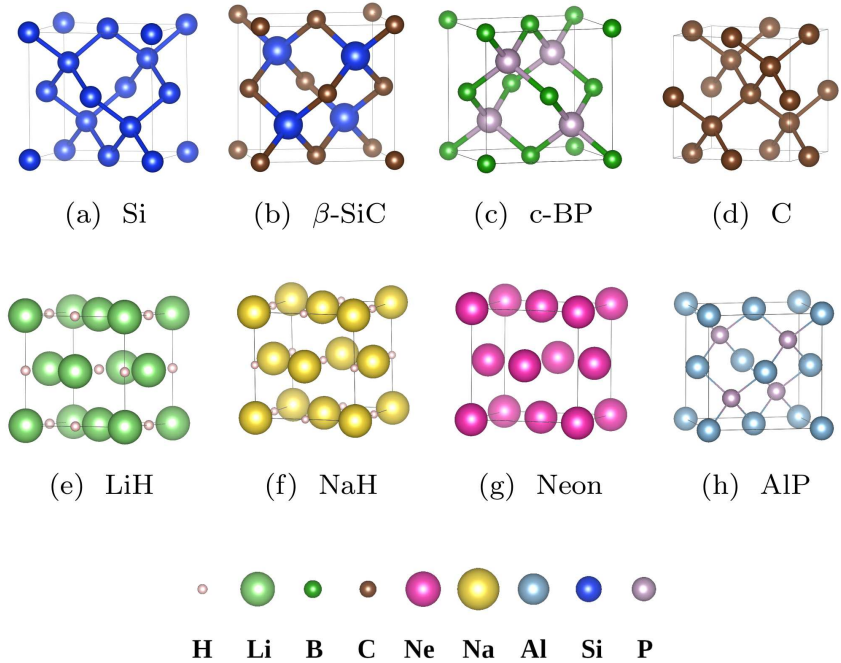}
	\caption{Crystal structures and unit cells of representative periodic systems (silicon (Si), beta silicon carbide ($\beta$-SiC), cubic boron phosphide (c-BP), diamond (C), lithium hydride (LiH), sodium hydride (NaH), solid neon (Ne), and aluminum phosphide (AlP)) studied in this work.}
    \label{fig:crystal}
\end{figure}

The periodic systems investigated in this work (\emph{cf.} Fig. \ref{fig:crystal}) crystallize in the cubic structure with the F$\bar{4}$3m (216) space group (c-BP, $\beta$-SiC, AlP), the Fd$\bar{3}$m (227) space group (Si, C), or the Fm$\bar{3}$m (225) space group (NaH, LiH, Ne). In these crystal structures, each atom is coordinated to four other atoms. The PBE optimized lattice parameters for the systems studied are summarized in the Table \ref{tab:modulus}. 

\begin{figure}[h!]
    \centering
    \includegraphics[width=0.5\columnwidth]{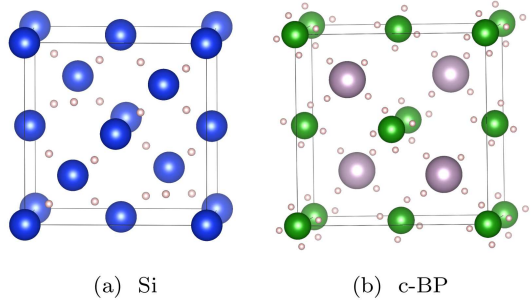}
	\caption{Final set of Wannier charge centers (shown as small circles) computed using our Wannier-Fermi-L\"{o}wdin approach for bulk silicon (Si) and cubic boron phosphide (c-BP).}
    \label{fig:WCC}
\end{figure}

A ground state reference PBE calculation was carried out to obtain Bloch states, which were further utilized to get Wannier functions.
Using Eqn. \ref{eqn:wcc}, we calculated the Wannier charge center positions by maximizing the localization of Wannier functions. 
The self-interaction corrected total energy per unit cell was calculated for this set of WCCs, which was subsequently minimized with respect to the WCC positions using a numerical conjugate gradient approach. The final set of WCCs along with the atoms in the unit cell for the case of silicon and cubic boron phosphide are as shown in Fig. \ref{fig:WCC}. The WCCs in these covalently-bonded systems lie along the line joining adjacent atoms, and their exact positions vary according to the electronegativity of the atoms. 


\fulltable{ 
\label{tab:modulus} Bulk moduli ($B$) of various solids calculated using the Wannier-Fermi-L\"{o}wdin function method. The computed PBE and WFL-PBE-SIC band gaps are listed with the reference HSE and experimental (Expt.) band gaps. }

\scriptsize
\begin{tabular*}{\textwidth}{@{}clcccccccc} 
\br
Sr. & System & Lattice (PBE) & $B$ (PBE) & $B$ (WFL-SIC) & Expt. $B$ & \multicolumn{4}{c}{Band gap (eV)} \\
No. &  & a (\AA{})  & GPa & GPa & GPa & PBE & HSE & WFL-SIC  & Expt.\\
\mr

1 & Si   & 5.475 & 89.4   & 96.8   & 98.8 {\cite{expt-bulk-modulus} } & 0.6 & 1.21 {\cite{hybrid-bandgaps} } & 1.7  & 1.17 {\cite{Kittel}}  \\ 
2 & c-BP & 4.538 & 161.8  & 179.0  & 174 {\cite{bulk-modulus-cBP}}  & 1.2 & 2.13 {\cite{hybrid-bandgaps} } & 2.5  & 2.40 {\cite{expt-bulk-modulus} }\\ 
3 & $\beta$-SiC & 4.348 & 210.8  & 225.0  & 227 {\cite{expt-bulk-modulus} }  & 1.3 & 2.32 {\cite{hybrid-bandgaps} } & 4.7  & 2.42 {\cite{expt-bulk-modulus} }\\ 
4 & C    & 3.567 & 439.0   & 422.3 & 442 {\cite{expt-bulk-modulus} }  & 4.1 & 5.43 {\cite{hybrid-bandgaps} }  & 5.3  & 5.47 {\cite{Kittel}}\\ 

5 & LiH  & 4.028 & 37.17  & 36.20  & 32.3 {\cite{expt-bulk-lih}}  & 3.0 & - & 3.6  & 4.99 {\cite{expt-gap-lih-nah} }\\ 
6 & NaH  & 4.833 & 23.13  & 22.9   & 19.4 {\cite{expt-bulk-nah}}  & 3.8 & - & 4.9  & 5.68 {\cite{expt-gap-lih-nah} }\\
7 & Ne   & 4.570 & 1.64   & 1.45   &  1.2 {\cite{expt-bulk-neon}}  & 11.6 & - & 21.9 & 21.58 {\cite{expt-gap-neon} }\\
8 & AlP  & 5.512 & 82.39  & 83.19  & 86.5 {\cite{expt-bulk-alp}}  & 1.6 & - &  4.0  & 2.45 {\cite{semi-mater} }\\
\\
\mr
\multicolumn{3}{r}{Mean Absolute Error}                 & 6.74   & 4.64   & --     & 2.37  & 0.11   & 0.76  &  -- \\
\multicolumn{3}{r}{Mean Absolute Relative Error}        & 0.125  & 0.08   & --     & 0.40  & 0.05   & 0.20  & -- \\
\br
\end{tabular*}
\endfulltable

\subsection{Bulk Modulus}

The bulk modulus ($B$) is an important mechanical property in the design and selection of materials\cite{expt-bulk-modulus, pbe-bulk-modulus} and is often used as a metric for benchmarking new electronic structure methods. Elastic properties such as the bulk modulus provide insight into the interatomic and bonding environments in these solid materials.\cite{bonding-bulk-modulus} Using the equilibrium volume, $V_0$, we calculated the bulk modulus, $B$, by computing the second derivative of a fitted energy curve with respect to volume, given by:
\begin{equation}
B = \frac{1}{V_0} \frac{\partial^2 E}{\partial V^2}.
\end{equation}
In Table \ref{tab:modulus}, we summarize the bulk modulus of various solids calculated using the PBE and WFL-self-interaction-corrected total energies. The self-interaction corrected total energy has two components, namely, Coulomb and XC contributions, and we take a $\frac{2}{3}$ fraction of each of these components, which is a standard procedure for scaling down the SIC over-correction.\cite{fredy-ravindra-fractional-sic} For comparison, we also list the experimental bulk modulus values for these solids in Table \ref{tab:modulus}. Finite-temperature and zero-point phonon effects were not included in our calculations.

As shown in Table \ref{tab:modulus}, for most of the systems, the WFL-SIC bulk moduli are an improvement over the PBE calculations and match more closely with the experimental benchmarks. It is worth noting that the bulk modulus is naturally dependent on the density of valence electrons $\rho$.\cite{highest-bulk-modulus} Since self-interaction corrections affect both the valence electron density and the total energy, these corrections would also manifest themselves in the bulk moduli of these materials. Indeed, Table \ref{tab:modulus} shows that the mean absolute error (MAE) and mean absolute relative error (MARE) for the WFL-SIC bulk moduli are 4.64 and 0.08 GPa, respectively, which are much smaller than their PBE counterparts.

\subsection{Electronic Structure}

In addition to the bulk moduli described previously, we also investigated the performance of the WFL-SIC approach for predicting electronic bandstructures of our periodic systems (as a side note, isolated molecular systems have flat dispersionless bands (independent of momentum, $k$), so the $\epsilon_\mathrm{HOMO}$ values summarized in Table \ref{tab:molecular} are already a performance check on the valence bands of those 9 molecules). Within the WFL-SIC formalism, the electronic bandstructures are calculated by solving the self-interaction corrected Hamiltonian:
\begin{equation}
H_{\mathrm{SIC}}^{\mathrm{WFL}} = H_{\mathrm{PBE}}^{0} + \hat{V}^{\mathrm{WFL-SIC}},
\end{equation}
where $H_{\mathrm{PBE}}^{0}$ is the PBE Hamiltonian and $\hat{V}^{\mathrm{WFL-SIC}}$ is the SIC potential computed using the WFL states. We investigated eight periodic systems, whose experimental band gaps span the range from 1.17 to 21.58 eV, which covers both insulators and semiconducting materials. In Fig. \ref{fig:bands} we plot the spectral function (which depicts the unfolded bandstructures) of all these materials. The SIC-corrected bandstructures are overlaid with the corresponding PBE results for comparison. 

\subsubsection{Electronic Band Gaps}
The computed PBE and WFL-SIC band gaps are summarized in Table \ref{tab:modulus} with the available reference HSE and experimental band gaps. From Fig. \ref{fig:bands}(a), we observe a significant opening of the $E_{\Gamma X}$ gap in the case of silicon, a small band gap material. Using our WFL-SIC approach, the  $E_{\Gamma X}$ gap increases to 1.7 eV for silicon, compared to the PBE value of 0.64 eV (the experimental value is 1.17 eV). An intermediate band gap material, $\beta$-SiC, exhibits a similar SIC overcorrection in which the 1.3 eV PBE band gap is widened to 4.7 eV (almost twice the experimental value). 

The WFL-SIC procedure for another intermediate band gap semiconductor material, cubic boron phosphide, gives a band gap of 2.5 eV, which is within one percent of the experimental value (2.4 ev) and also outperforms the HSE result (2.13 eV).\cite{hybrid-bandgaps,koopman-marzari} This is a significant improvement over the PBE result of 1.2 eV, which severely underestimates the experimental benchmark. Similarly, the band gap of diamond, a wide band gap insulator, shows a significant improvement in which the PBE band gap of 4.10 eV is increased to 5.3 eV, which matches well with the experimental value of 5.47 eV. 

The PBE band gap for the metal hydrides with a rocksalt crystal structure are substantially corrected using the WFL-SIC procedure. Lithium hydride exhibits a 20 \% increase in the band gap, pushing it to 3.6 eV. The indirect band gap in   sodium hydride, on the other hand, shows a 1.1 eV gap opening. Aluminum phosphide, which has the same space group as $\beta$-SiC, shows a similar overcorrection. The 1.6 eV PBE gap is increased to 4 eV compared to the 2.45 eV experimental value.\cite{semi-mater} Our calculations for bulk neon, on the other hand, show an excellent band gap correction in which the 11.6 eV PBE band gap is corrected to 21.9 eV with WFL-SIC, which closely matches the experimental value of 21.58 eV. \cite{expt-gap-neon}

In conjunction with the electronic property predictions of molecular systems in Table \ref{tab:molecular}, the predictions using the WFL-SIC approach are relatively good and an improvement over the semi-local PBE results.

\subsubsection{Electronic Bandstructures}

\begin{figure*} 
    \centering
    \includegraphics[width=\columnwidth]{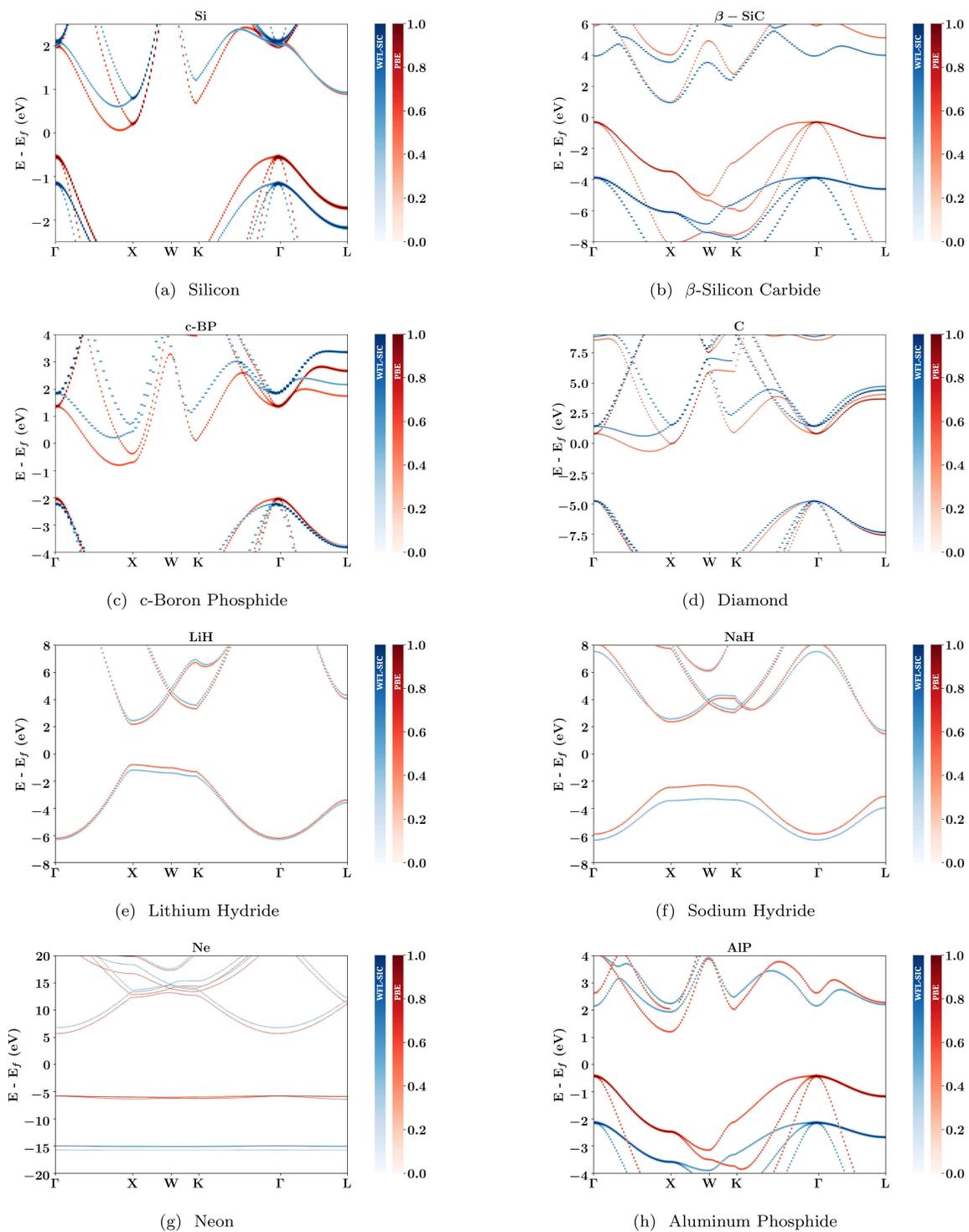}
    \caption{Wannier-Fermi-L\"{o}wdin SIC-corrected (blue dots) and PBE (red dots) bandstructures of various periodic systems examined in this work. The colormap denotes the normalized values of the spectral function of the unfolded bands.}
    \label{fig:bands}%
\end{figure*}

Upon closer inspection of the various band structures computed in this work, we find that the valence band dispersion predicted by WFL-SIC and PBE for silicon (\emph{cf.} Fig. \ref{fig:bands}(a)) is similar near the $\Gamma$ point. However, the WFL-SIC valence band maximum (VBM) at $\Gamma_{25^{'}}$ is shifted downwards by $\sim$ 0.6 eV, while the WFL-SIC bands near the conduction band minimum (CBM) at the X$_{1}$ point are shifted upwards by the same amount. In addition, the dispersion of the WFL-SIC conduction bands deviates slightly from the PBE results.  In the case of $\beta$-SiC (\emph{cf.} Fig. \ref{fig:bands}(b)), a major downshift of the valence bands is observed in the WFL-SIC calculations. Although the band dispersion remains nearly identical, the WFL-SIC VBM at $\Gamma_{25^{'}}$ is pushed down by 3.5 eV, while the CBM at the X$_{1}$ point remains unchanged. 

In contrast to the aforementioned materials, the WFL-SIC bandstructure of cubic boron phosphide shows a significant improvement over the PBE calculations (\emph{cf.} Fig. \ref{fig:bands}(c)). While the valence bands remain almost identical to those of PBE, the WFL-SIC conduction bands are shifted upwards, keeping the dispersion intact. The band structure of diamond (\emph{cf.} Fig. \ref{fig:bands}(d)), a wide band gap insulator, undergoes a similar transformation when the WFL-SIC formalism is applied. The band dispersion is almost preserved for the valence bands, but  a slight flattening is observed for the WFL-SIC conduction band near the X point. 

In the case of LiH, the WFL-SIC corrected bandstructure (\emph{cf.} Fig. \ref{fig:bands}(e)) shows a gap opening near the $X$ point. The band dispersion remains nearly the same compared to the PBE bandstructure. Sodium hydride, on the other hand, exhibits a substantial downward shift of valence bands (\emph{cf.} Fig. \ref{fig:bands}(f)). The band dispersion remains nearly unchanged in this case also. 
Bulk neon, which is an extreme insulator, shows a dramatic correction of bands (\emph{cf.} Fig. \ref{fig:bands}(g)). In particular, the closely-spaced valence bands at around -6 eV are pushed downwards by a significant amount, correcting the PBE bandstructure to closely match the experimental one. 
Finally, aluminum phosphide, whose bandgap is severely underestimated by PBE, exhibits a significant correction using the WFL-SIC method (\emph{cf.} Fig. \ref{fig:bands}(h)). The valence bands are lowered by 1.6 eV, and the dispersion of the conduction bands is also modified. A nexus point in the conduction bands (along $\Gamma$-X) is lifted by the WFL-SIC method.

Upon closer examination of the results for diamond and bulk neon, we obtain an extremely close exact match with the experimental benchmarks, whereas the results for $\beta$-SiC and AlP show larger deviations. Both diamond and neon are wide band gap insulators; hence, the Wannier functions are very well localized. This, in turn, gives the best estimate of Wannier charge centers (a$_m$), which are crucial parameters for obtaining the WFL-SIC corrections. 
For the rest of the materials (except c-BP), we observe a larger deviation from the experimental benchmarks. We associate this deviation to the accuracy of Wannier charge center positions. The Wannier spread (which is proportional to $\langle \tilde{r_n}^2 \rangle$) is comparatively larger for small band gap materials, which contributes to the error bar in the SIC potential and is parametrically dependent on these centers. Our WFL-SIC approach, when applied to c-BP, diamond, and neon, gives results that match almost perfectly with the experimental measurements. It is important to note that a small Wannier spread (or precise locations of Wannier charge center positions) results in a better Wannier-Fermi-Lowdin function. In the case of insulators, the Wannier functions are exponentially localized\cite{exponential-wannier} and, therefore, the Wannier spread is minimal for such materials. As such, our method is expected to perform better for predicting the band structure of these materials.

Finally, to assess and compare the computational cost of our WFL-SIC approach, we performed HSE06 and quasiparticle-based G$_{0}$W$_{0}$ calculations for silicon as a prototypical benchmark case. To perform this computational timing test, 16-core Intel Xeon CPUs (E5-2640 v3) clocked at 2.60 GHz were used to evaluate eigenvalues across 200 kpoints in the BZ. A G$_{0}$W$_{0}$ calculation was performed using a 300 eV plane-wave cutoff in conjunction with the plasmon-pole approximation as implemented in GPAW.  Upon convergence, the WFL-SIC, HSE, and G$_{0}$W$_{0}$ calculations took 16.2, 32.16, and 656 CPU-hours to complete, respectively. As such, these computational timing tests indicate that the WFL-SIC approach is twice as fast as Hartree-exchange-based hybrid density functionals and significantly more efficient than the computationally expensive quasiparticle-based G$_{0}$W$_{0}$ methods.

\section{Conclusions}
In summary, we have provided the first formulation and implementation of a Wannier-Fermi-L\"owdin approach for the efficient computation of self-interaction corrections for DFT calculations of periodic systems. This computational approach is carried out by minimizing the SIC energy by varying an energy functional with respect to the Wannier charge centers of the periodic system. In particular, this functional minimization problem involves only $3N$ parameters compared to the conventional $N^2$ parameters used in a full SIC calculation, resulting in substantial computational savings. To test our implementation, we have benchmarked our approach across 17 prototypical molecular solids, semiconductors, and wide-bandgap materials with different crystal structures that span a wide range of electronic properties. Our results indicate that our WFL-SIC approach partially removes the spurious self-interaction errors in molecular as well as periodic systems to give better ionization potentials, band gaps, and bulk moduli compared to those predicted by semilocal functionals. In our final test of our method, we also showed that the WFL-SIC approach is computationally more efficient than either the HSE hybrid functional or the quasiparticle-based G$_{0}$W$_{0}$ method (while still showing improved accuracy over the PBE results). As such, our WFL-SIC approach could be a viable option for obtaining improved electronic properties for massive periodic systems where HSE (or G$_{0}$W$_{0}$) calculations are prohibitively out of reach. 

\ack
This work was supported by the U.S. Department of Energy, Office of Science, Early Career Research Program under Award No. DE-SC0016269. Helpful conversations with Dr. Fredy W. Aquino are greatly acknowledged.

\bibliographystyle{unsrt}
\bibliography{apssamp}
\end{document}